\documentclass[twocolumn]{aastex63}

\usepackage[scaled]{helvet}

\usepackage[utf8]{inputenc}
\usepackage[T1]{fontenc}

\submitjournal{ApJL}

\shorttitle{Super-Earth redox hysteresis}
\shortauthors{Lichtenberg}

\graphicspath{{./}{figures/}}

\begin{document}

\title{Redox hysteresis of super-Earth exoplanets from magma ocean circulation}

% \correspondingauthor{}
\email{tim.lichtenberg@physics.ox.ac.uk}

\author[0000-0002-3286-7683]{Tim Lichtenberg}
\affiliation{Atmospheric, Oceanic and Planetary Physics, Department of Physics, University of Oxford, UK}

\begin{abstract}
Internal redox reactions may irreversibly alter the mantle composition and volatile inventory of terrestrial and super-Earth exoplanets and affect the prospects for atmospheric observations. The global efficacy of these mechanisms, however, hinges on the transfer of reduced iron from the molten silicate mantle to the metal core. Scaling analysis indicates that turbulent diffusion in the internal magma oceans of sub-Neptunes can kinetically entrain liquid iron droplets and quench core formation. This suggests that the chemical equilibration between core, mantle, and atmosphere may be energetically limited by convective overturn in the magma flow. Hence, molten super-Earths possibly retain a compositional memory of their accretion path. Redox control by magma ocean circulation is positively correlated with planetary heat flow, internal gravity, and planet size. The presence and speciation of remanent atmospheres, surface mineralogy, and core mass fraction of atmosphere-stripped exoplanets may thus constrain magma ocean dynamics.
\end{abstract}

\keywords{Extrasolar rocky planets, Super Earths, Planetary interior, Exoplanet atmospheres, Exoplanet surfaces}

\section{Volatiles in and on rocky planets} \label{sec:intro}

The volatile inventory governs long-term climate and tectonics of terrestrial worlds, but their inheritance and distribution between core, mantle, and atmosphere are not well understood. The apparent radius transition from sub-Neptune to super-Earth exoplanets \citep{2017AJ....154..109F} has illustrated a previously unappreciated path to form rocky worlds via loss of their nebula-derived gaseous envelopes \citep{Bean21}, but the presence and possible composition of remanent atmospheres on the envelope-stripped planet population remains in doubt. High mean molecular species, such as water or carbon compounds, may be sourced from thermallly pristine materials from wide heliocentric orbits \citep{Venturini20}, but Solar System petrochronology indicates accretion of the terrestrial planets from devolatilized building blocks \citep{Lichtenberg19a,Lichtenberg21Science}. Exogenous incorporation of water-rich solids into the planet during accretion oxidizes the planet and leads to higher FeO abundances in the mantle \citep{ET08}, which affects the atmospheric composition that can be sustained in equilibrium with the planetary interior on terrestrial worlds \citep{Gaillard21}. 

This suggests that after loss of primordial hydrogen-rich atmospheres, the composition of outgassed secondary atmospheres of rocky planets is governed by the rate and chronology of volatile addition during accretion. However, endogenous processes operating in the interior of the planet may overprint the primitive assemblage inherited from formation. These can be particularly efficient while the rocky interiors of young planets are molten due to accretional heat \citep{Ikoma18,Chao20}, which may be sustained for billions of years on sub-Neptune exoplanets \citep{Vazan18,Kite20a}. 

One mechanism that can alter the primordial mantle composition is the disproportionation of iron species from ferrous to ferric iron ($3\mathrm{Fe}^{2+} \rightarrow \mathrm{Fe}^{0} + 2\mathrm{Fe}^{3+}$) at high pressures \citep{Frost04,Wade05}, which can increase the oxidation state of the lower planetary mantle. This has been suggested to equilibrate the chemical composition of mantle and overlying atmosphere via convectional mixing between deep and shallower parts of the Earth and affect the chemical speciation of the outgassed atmosphere \citep{2012EPSL.341...48H,Armstrong19}. 

Additionally, the simultaneous presence of FeO mineral phases in magma oceans and primitive hydrogen-rich atmospheres has been suggested to generate water in the interiors of terrestrial planets \citep{Ikoma06} and sub-Neptune exoplanets \citep{KS21} via the reaction $\mathrm{FeO} + \mathrm{H}_{2} \rightarrow \mathrm{Fe}^0 + \mathrm{H}_{2}\mathrm{O}$. For this mechanism, the conversion efficiency to H$_2$O depends on the rate of H$_2$ admixture from the atmosphere into the planetary interior. Similar to the disproportionation mechanism, this may require mixing between near-surface, H$_2$-satured and deeper, unsaturated layers of the magma ocean \citep{Olson19} to homogenize the planetary interior. Therefore, convectional mixing is crucial for establishing chemical equilibration between mantle and atmosphere as a result of iron disproportionation or endogenous water production during magma ocean stages. 

For net oxidation of the mantle-atmosphere system, both redox reactions additionally require removal of reduced Fe$^0$ from the silicate mantle to the metal core. However, while crucial for the movement of volatiles between atmosphere and mantle, eddy diffusion in turbulent flows can carry enough kinetic energy to entrain particles of a different substance. In highly turbulent regimes with high fluid velocities, the internal stirring in magma oceans can curb the transfer of liquid iron droplets from the mantle to the metal core \citep{Solomatov93a}, and thus alter the degree of chemical equilibration between atmosphere, mantle, and core in terrestrial planets \citep{keppler19}. 

Particle entrainment may be particularly relevant for the magma oceans sustained in the interiors of sub-Neptune exoplanets. These are deeper than those on terrestrial planets and operate under higher gravity, which alters their internal flow and cooling regime \citep{Vazan18}. It is hence unclear if both volatile equilibration between atmosphere and mantle and iron equilibration between mantle and core can simultaneously operate with high efficacy. 

Here, I address this issue in the regime that governs the sub-Neptune to super-Earth transition and investigate whether the internal magma circulation can in principle be a regulating factor for the chemical equilibration between mantle and core. 
\begin{figure*}[tbh]
\centering
\includegraphics[width=.49\textwidth]{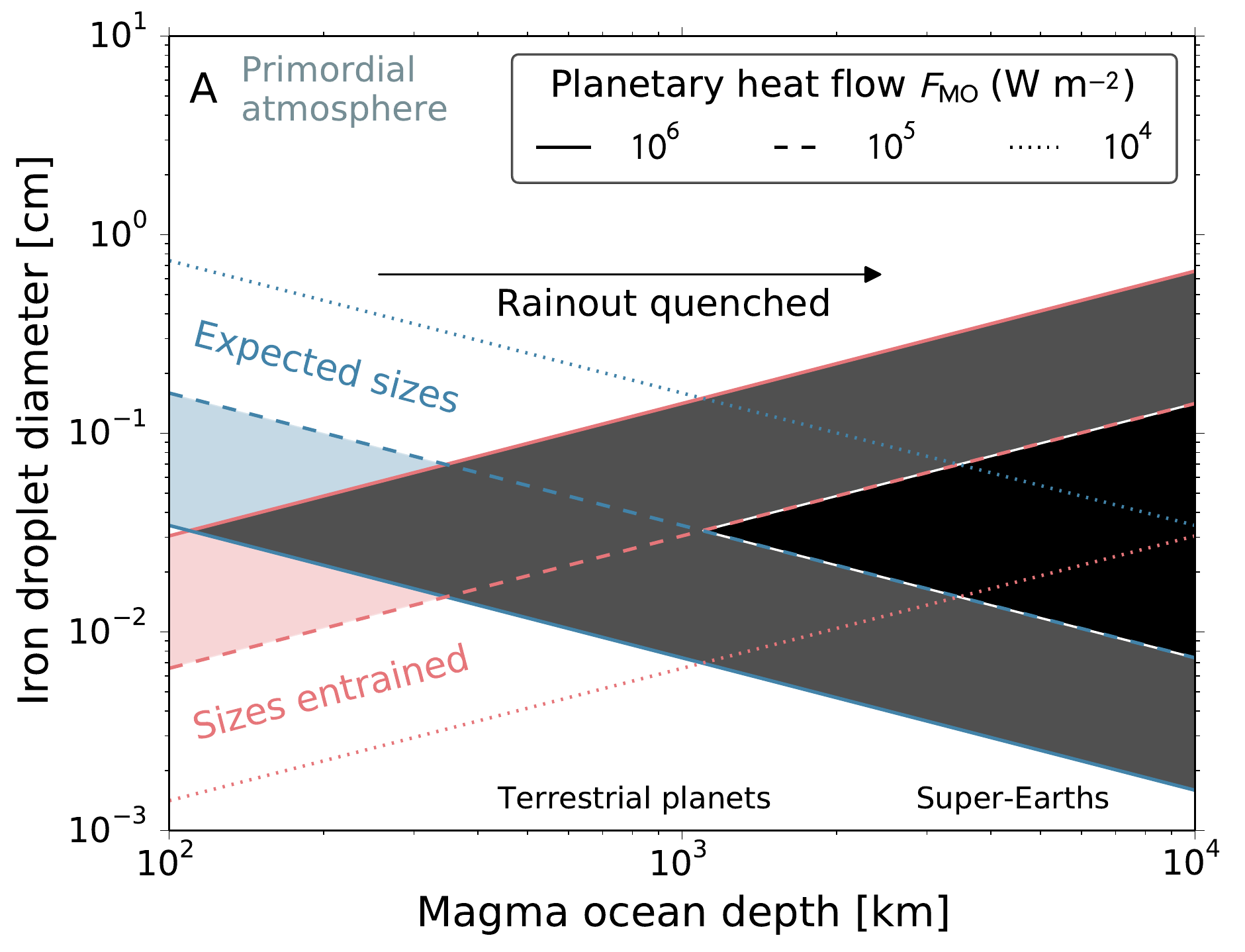}
\includegraphics[width=.49\textwidth]{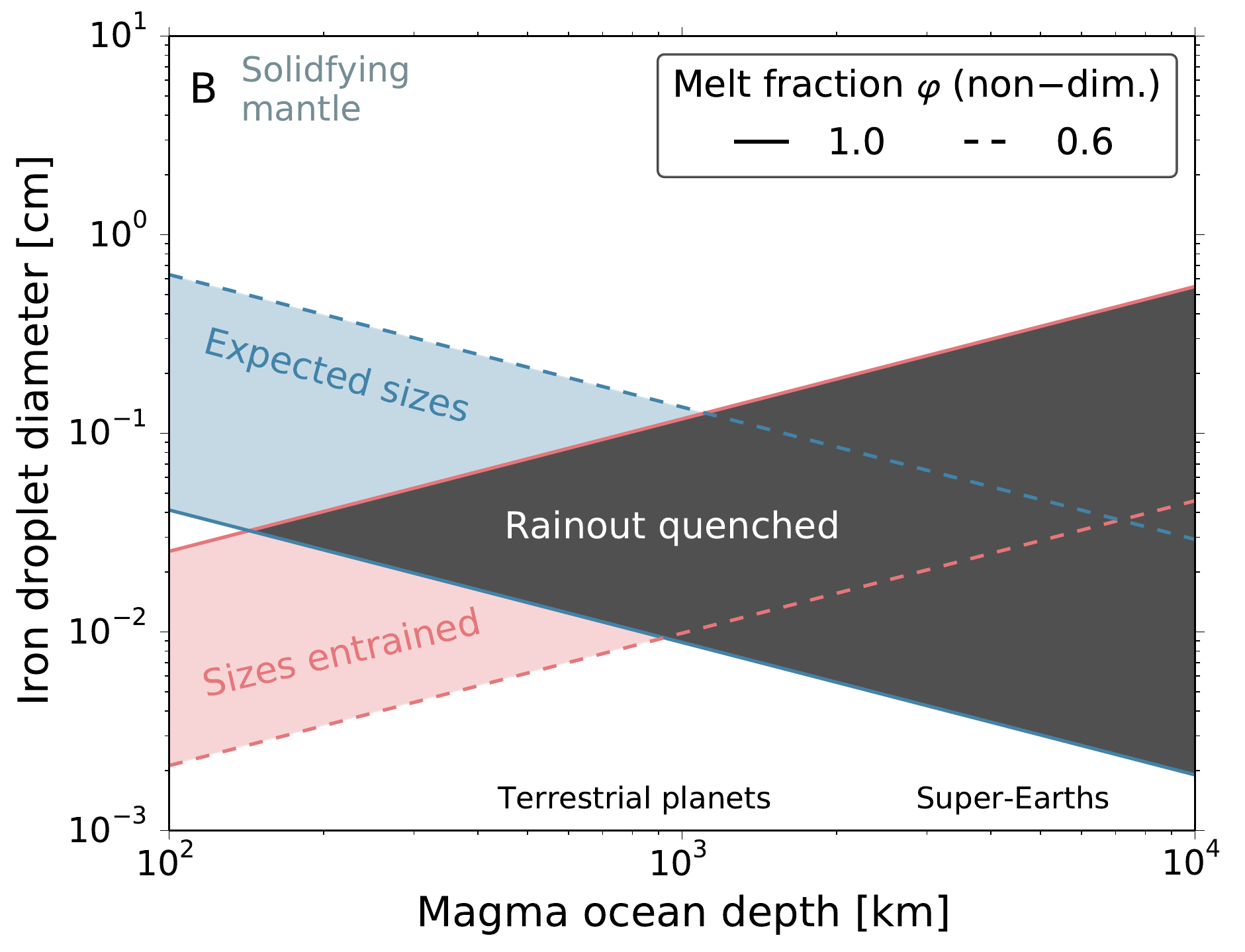}
\caption{\textsf{Competition between liquid iron droplet rainout and entrainment in magma oceans. \textbf{(A)} Case for a fixed heat flux that is set by a nebula-derived, primordial atmosphere. Upper size limit for iron droplets in blue ('Expected sizes') versus droplet sizes that can be entrained in the magma flow in red ('Sizes entrained'). The various lines show planetary heat flows of $F_\mathrm{MO} = 10^6$ W m$^{-2}$ (solid), $10^5$ W m$^{-2}$ (dashed), and $10^4$ W m$^{-2}$ (dotted). Right of where red and blue cross each other (dark gray and black shaded, respectively), droplets may be suspended in the large-scale circulation of the magma ocean and escape rainout onto the iron core ('Rainout quenched'). \textbf{(B)} Same as A but for the case of a solidifying magma ocean with decreasing silicate melt fraction $\varphi$ from 1.0 (solid lines) to 0.6 (dashed lines). See text for discussion.}}
\label{fig:entrainment}
\end{figure*}
\section{Turbulent flow impedes iron rainout} \label{sec:results}
Reduced iron can be generated by the redox reactions introduced above throughout the molten planetary mantle. Whether these liquid iron droplets are suspended or rain out from the magma ocean and merge with the iron core is governed by the competition between gravity (denser particles sink) and kinetic energy in the magma flow (turbulent eddies entrain particles). The flow regime in buoyancy-driven convection can be described by the Rayleigh number
\begin{equation}
\mathrm{Ra} = \alpha \rho g  \Delta T D^{3}/(\kappa \eta),
\end{equation} 
with thermal expansivity $\alpha$, density of the liquid magma $\rho$, gravity acceleration in sub-Neptunes $g \approx 30$ m s$^{-2}$, superadiabatic temperature gradient of the planet $\Delta T$, depth of the magma column $D$, thermal diffusivity $\kappa$, and melt viscosity $\eta$. The associated heat flux in the soft turbulence limit is \citep{solomatov15}
\begin{equation}
F_\mathrm{MO} = 0.089 \, k \, \Delta T \, \mathrm{Ra}^{1/3}/D, \label{eq:heatflux}
\end{equation}
with thermal conductivity $k$. 

Liquid iron droplets in the magma converge toward a characteristic size due to the competition between the stagnation pressure and internal pressure caused by surface tension \citep[][]{Rubie03}, 
\begin{equation}
d_\mathrm{droplet} \approx \frac{\sigma \cdot \mathrm{We}}{\left(\rho_{\mathrm{Fe}}-\rho\right) v_{\mathrm{MO}}^{2}},
\end{equation}
with the surface energy of the iron droplet-magma interface $\sigma$, Weber number $\mathrm{We} \approx 10$, and density of liquid iron droplets $\rho_{\mathrm{Fe}}$. Convective velocities in this regime scale as \citep{solomatov15}
\begin{equation}
v_{\mathrm{MO}} \approx 0.6\left(\frac{\alpha g l F_\mathrm{MO}}{\rho c_{\mathrm{p}}}\right)^{1 / 3},
\end{equation}
with mixing length $l \sim D$ and heat capacity $c_{\mathrm{p}}$. Relations between melt density and mantle potential temperature relate the convective scalings to numerical closures for fully and partially molten magma oceans defined by globally averaged mantle potential temperatures and melt fractions, as given in \citet[][Tab.~1]{Lichtenberg18}.

At very high Rayleigh number, the stresses associated with turbulence in the fluid flow increase substantially \citep{Shraiman90}. The associated empirical limit to flow suspension for settling particles is \citep{solomatov15}
\begin{equation}
d_\mathrm{crit} \lesssim \frac{\rho\left(v_{\mathrm{MO}} / 60\right)^{2}}{0.1\left(\rho_{\mathrm{Fe}}-\rho\right) g}.
\end{equation}
Particles larger than this critical size are expected to rain out from the magma ocean and enter the core. Particles smaller than this size threshold can potentially be suspended or reeintrained by the large scale circulation of the magma flow in the interior of the planet.

Fig.~\ref{fig:entrainment} explores this competition between iron droplet entrainment by the magma circulation and settling for \emph{(i)} the steady-state case of a fully-molten magma ocean with a fixed heat flux, and \emph{(ii)} the case of a solidifying interior with decreasing melt fraction. The steady-state case \emph{(i)} (\emph{'Primordial atmosphere'}), displayed in Fig.~\ref{fig:entrainment}A, assumes that the heat flux of the magma ocean is controlled by the nebula-derived primordial atmosphere. The associated heat flux can vary by orders of magnitude, depending on the total mass of the atmosphere, its temperature-pressure structure, and the uncertain opacity of H-He-dominated gas at $T \gtrsim 2000$ K \citep{Vazan20b}. In the \emph{Primordial atmosphere} scenario, iron droplets can rain out onto the core for magma ocean depths $\lesssim$100 km and $\lesssim$1000 km for $10^6$ W m$^{-2}$ and $10^5$ W m$^{-2}$, respectively (Fig.~\ref{fig:entrainment}A). For deeper magma oceans at these heat fluxes, such as those in sub-Neptunes, the kinetic energy carried by the flow can potentially suspend iron droplets and halt core formation (gray and black areas). For heat flows $F_\mathrm{MO} \lesssim 10^4$ W m$^{-2}$ droplet suspension cannot be achieved in the sub-Neptune regime because the kinetic energy of the convection cells decreases substantially. 

Fig.~\ref{fig:entrainment}B illustrates case \emph{(ii)} (\emph{'Solidifying mantle'}) for a crystallizing magma ocean. The surface temperature is fixed close to the rheological transition of peridotite melt-solid aggregates at 1500 K and the heat flux is self-consistently calculated via Eq. \ref{eq:heatflux} by scaling the mantle potential temperature with the global melt fraction \citep{Lichtenberg18}. Below the rheological transition at silicate melt fraction $\varphi \sim 0.5$ \citep[][]{Costa09}, magma viscosity increases rapidly, which slows down convective motions and the mantle starts behaving rheologically like a solid rather than a liquid. Thus, I vary melt fraction between 1.0 to 0.6, where the mantle is in the fluid-like flow regime, and the above scalings are applicable. Fig.~\ref{fig:entrainment}B suggests that crystallizing sub-Neptune interiors deeper than $\gtrsim$ 200 km can in principle entrain liquid iron droplets. This increases substantially with the onset of solidification, such that only magma oceans deeper than $\gtrsim$ 7000 km can entrain iron droplets when the mantle melt fraction approaches the rheological transition (where the dashed lines cross in Fig.~\ref{fig:entrainment}B) because the convective velocities in the magma ocean decrease with increasing solid fraction and thus increasing melt viscosity.

These results illustrate that the circulation regime of the magma ocean may influence the chemical equilibration between core and mantle. This means that during the sub-Neptune to super-Earth transition, reduced iron, which may be generated by the interior redox reactions introduced above, can be retained in the magma together with produced water and perovskite, where it can in principle establish chemical equilibrium in the melt phase.

\section{Discussion} \label{sec:discussion}

\subsection{Uncertainties}

The relations employed in this work quantify buoyancy-driven flow based on Rayleigh–B{\'e}nard convection. Sub-Neptunes may not be in this regime because the strong blanketing from primitive H$_2$ atmospheres can efficiently protract heat loss \citep{Vazan18,KiteBarnett2020PNAS,Lichtenberg21JGRP}. However, if cooling in sub-Neptunes is sufficiently slow the flow will transition to laminar and eventually a conduction-dominated, layered structure. If this were the case, then the chemical equilibration between atmosphere and mantle would be in question, which similarly would inhibit redox reactions involving envelope-derived hydrogen. This would mean that internal water production may be limited to a transition flow regime, where convective stirring is strong enough to draw down enough H$_2$ into the liquid mantle, but not strong enough to prevent iron droplet rainout. For iron disproportionation this would imply that oxidation of the upper mantle of super-Earths is quenched at values set by the material composition acquired during accretion.

Generally speaking, the results presented here suggest that there is a converse relation between homogenization of the interior and composition stasis due to very strong convection in highly turbulent magma oceans. The former is required for the transfer of reducing power between mantle and core (in the form of iron droplets), and equilibration between mantle and atmosphere (ingassing and outgassing of volatiles), but at extreme Rayleigh numbers the convection in the magma flow may entrain iron droplets, which quenches these connections between the sub-systems of sub-Neptune exoplanets. As illustrated in Fig.~\ref{fig:entrainment}, there is a sensitive connection between heat flux, melt fraction, and entrainment. Observed exoplanets, however, are typically billions of years old. If these worlds can sustain a fully liquid magma ocean in their interiors during this time, their average heat flux must be lower than the ones for which the rainout quenched regime is stable. This means that during solidification the initial rainout quenched regime may transition to calmer circulation regimes, where entrainment is less likely, and hence mantle-core differentiation can be achieved. 

However, because of the uncertainties related to the equation of state at high pressures this question directly connects the equilibration between core and atmosphere to the mode of mantle crystallization. For instance, middle-out crystallization \citep{Stixrude14} or spatially inhomogeneous redox state in the interior of sub-Neptunes complicate simple bottom-up crystallization scenarios, and may require spatially resolved models \citep{Bower19} to capture inhomogenous crystallization and redox evolution. If rainout is delayed until late magma ocean stages, the dilute suspension limit applied here may not be appropriate anymore and chemical fractionation may be governed by crystal-driven convection \citep{Culha20}.

Along the same route, the rainout quenched regime advocated here does not guarantee no mantle-core equilibration at all. Instead, evolution of the crystal size distribution, stochastic motion in turbulent eddies, and diffusion processes may continuously transfer suspended particles to the pre-existing metal core \citep{Ichikawa10,Patocka20}. Therefore, a more complete assessment of the rainout quenched scenario will require time- and spatially-resolved models of magma oceans to relate interior evolution to atmospheric properties \citep{Lichtenberg21JGRP}.

The composition of rocky exoplanet interiors may substantially differ from our Solar System and hence affect material properties. However, these affect the convection regime at a similar scale as the superadiabatic gradient, and are negligible in comparison with the cubic dependence of the Rayleigh number on magma ocean depth. If sub-Neptune interiors can cool and hence flow (or vice versa), their internal magma circulation is likely in a turbulent regime. Another complication of varying compositions is whether silicate-dominated magma oceans are formed at all. The presence of high FeO content in primitive planetary mantles may depend foremost on the accretion of oxidized, and hence water-rich solids \citep{ET08}. At high water-to-rock ratios the melting and degassing properties may differ substantially from terrestrial compositions \citep{Vazan20b}. If sub-Neptunes are formed predominantly beyond the water snow line and as a result the internal layering of silicate and volatile ice phases becomes more diffuse \citep{Vazan20} the scalings employed above may not be appropriate.

The model presented here treats the flow regime in a conservative limit, assuming that flow velocities scale akin to soft turbulence. However, for very deep magma oceans and hence very high Rayleigh numbers, hard turbulence becomes increasingly likely. In the hard turbulence regime convective motions may self-organise into large-scale circulation patterns, which increases the convective velocities relative to the soft turbulence regime \citep{solomatov15}. This would strengthen the conclusions of this manuscript. Transitions between soft and hard turbulent convective regimes are debated though.

Planetary rotation may affect the inertial forces in the magma flow and hence the settling behaviour of iron droplets. For example, \citet{Moeller13} find complex transitions in particle entrainment with increasing rotation rate. \citet{solomatov15} argue for a decreasing influence of rotation with increasing Ra, which would render rotation of minor importance for super-Earths relative to terrestrial-type magma oceans. Currently neither laboratory nor computational experiments \citep{Patocka20} can reach appropriate Rayleigh numbers to settle this question in the super-Earth regime.

Atmosphere-stripped, tidally-locked super-Earths may undergo hemispheric flow patterns \citep{Meier21}, where the direction of gravity in the interior is misaligned with the dominating temperature gradient and fluid currents. In this case the planet is partly solidified, and atmospheric hydrogen is already depleted, hence endogenic water production is not possible anymore. However, future observations of volcanic activity may offer possibilities to probe mantle composition \citep{Quick20} on such planets.  

\subsection{Implications \& observability}
\begin{figure*}[tb]
\centering
\includegraphics[width=.65\textwidth]{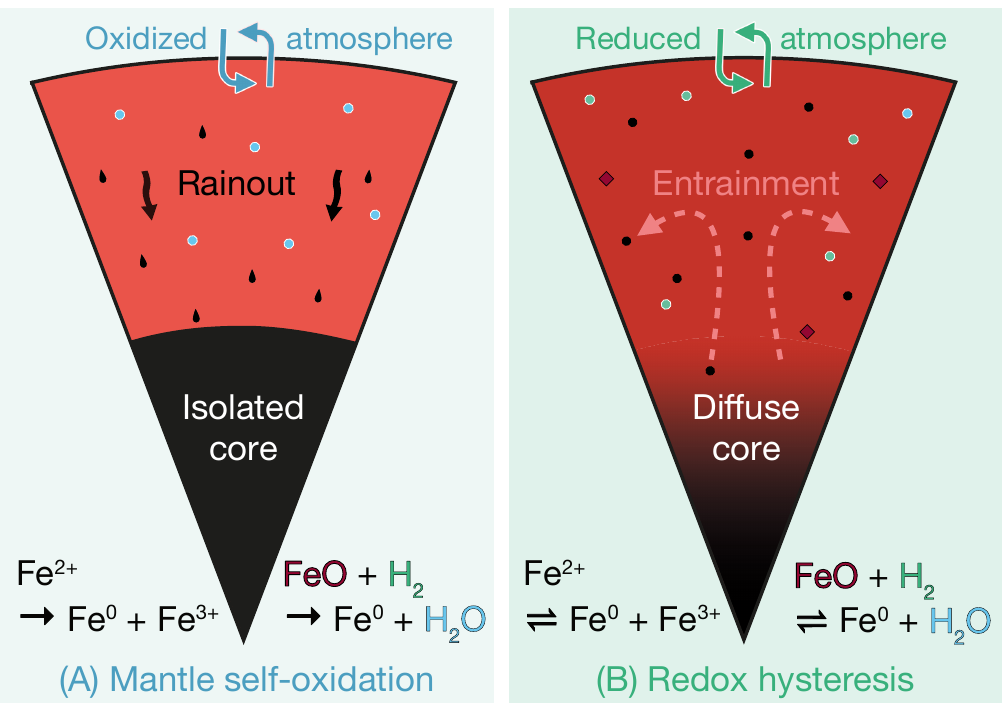}
\caption{\textsf{Illustration of the two possible end-member cases of magma ocean circulation with varying efficacy of redox reactions that drive changes in primordial mantle composition. \textbf{(A)} Magma ocean circulation is sufficiently vigorous to equilibrate magma and overlying atmosphere, but not energetic enough to entrain liquid iron droplets. These rain out onto the metal core, which reslts in net oxidation the mantle via iron disproportionation at high pressure and may sustain endogenic production of water. Secondary atmospheres generated by outgassing from such planetary mantles would be dominated by oxidized species, such as H$_2$O, CO$_2$, or SO$_2$. \textbf{(B)} In the rainout quenched regime magma circulation is highly turbulent such that the kinetic energy of the large-scale flow entrains liquid iron droplets. This regime may preserve the mantle composition inherited from accretion, diffuse the physical differentiation between silicate mantle and metal core, and result in secondary atmospheres dominated by reduced species, such as H$_2$, CO, or CH$_4$, and their photochemical derivatives.}}
\label{fig:illustration}
\end{figure*}
Whether redox reactions in the interiors of sub-Neptunes can drive compositional change alters the prospects for obervations of atmospheres on hot rocky exoplanets. If the mantle composition in super-Earths can sustain sufficiently reducing compositions, then atmospheres in equilibrium with internal magma oceans may be dominated by reduced compounds, such as H$_2$, CO, and CH$_4$ \citep{Gaillard21,Sossi20}, and their photochemical derivatives. Contrary to this, if the rainout quenched regime cannot be sustained and redox reactions irreversibly alter mantle composition, secondary atmospheres on super-Earths should be oxidized, dominated by species such as H$_2$O, CO$_2$, or SO$_2$. The former, reduced case would enhance the chances for detection of outgassed atmospheres with the James Webb Space Telescope and ground-based high-resolution spectroscopy. At the moment, the presence and speciation of atmospheres on hot rocky exoplanets is contentious \citep{Swain21,mugnai2021} due to the limitations of current astronomical observatories. Detailed characterization of individual, close-by transiting rocky planets may be the most direct path to investigate this question \citep{Trifonov21}, but population studies of atmospheric properties will be needed to address the statistical significance for planetary bulk compositions \citep{Wang19,Helled21}.

From a formation perspective, finding evidence for or against steam atmospheres on close-in super-Earths can further constrain the accretion path and composition of the atmosphere-stripped part of the Kepler radius valley \citep{RogersOwen21}. Finding evidence for H$_2$O on hot rocky exoplanets could be interpreted in two directions. \emph{(i)} Super-Earths inherit oxidized solids during accretion to either build an inventory of FeO that can later react with the H$_2$ atmosphere, or directly incorporate water from the accreted solids. \emph{(ii)} If the water can only be produced endogenously (no admixing of outer, water-rich materials) the magma oceans of such sub-Neptunes must be in a circulation regime that allows H$_2$ admixture and permits Fe$^0$ rainout. This is different to the buoyancy-driven flow regime expected for the terrestrial magma ocean as illustrated in this work. In addition, FeO would need to become incorporated into the planet's mantle in a reduced environment, suggesting mantle self-oxidation by disproportionation to be effective in sub-Neptune interiors.

The two different end-member scenarios that are suggested by this work are shown in Fig.~\ref{fig:illustration}. In case A the stirring velocities in the magma ocean are low enough for iron droplets to settle, which drives iron disproportionation and water generation by FeO reduction. In case B particles are entrained. In this case iron disproportionation and endogenic water production will be quenched at local equilibrium. In addition to limiting redox reactions, this preserves a larger mass of the H$_2$ that is partitioned into the magma ocean to outgas upon depletion of the atmospheric reservoir. This may prolong the presence of hydrogen-rich atmospheres on rocky exoplanets in addition to preserving the chemical composition of the mantle.

Steam atmospheres are harder to strip from a planet due to the high partition coefficient of water in magma and the increased blanketing effect of water relative to hydrogen species. This leads to prolonged and higher weight atmospheres on planets with endogenic water production. This predicts longer lifetimes for water-dominated magma oceans, higher mean molecular weights of remanent atmospheres, and lower scale heights. In the other case, H$_2$ atmospheres are easier to strip, partition less strongly into the magma, and insulate planets less for a given relative atmospheric mass fraction. Hence, all else being equal, such magma oceans live shorter lives and their atmospheres are larger due to lower atmospheric weight. The direct dependence on atmospheric heat flow and magma ocean depth suggests that planets with larger size and at wider heliocentric orbit (due to lower instellation) are more prone to fall in the rainout quenched regime.

In addition to bulk and atmospheric compositions, bare surface spectra of ultimately desiccated rocky planets \citep{Hu12,Koll19} or direct imaging of magma ocean planets at wider orbital separation \citep{Bonati19} may discriminate between magma ocean circulation regimes that enable internal water production and those that do not. The former predicts oxidized surfaces \citep{KS21}. The latter predicts that desiccated super-Earths feature basaltic or ultramafic surface mineralogy. Whether these surface alterations can be sustained and eventually observed depends on the degree of mantle overturn upon crystallization \citep{Tikoo17}, the possible development of basal magma oceans in super-Earth interiors \citep{Stixrude14}, and the thickness of the planetary crust that develops by magmatism on rocky planets \citep{Dyck21}.

\section{Summary \& conclusions} \label{sec:summary}
Internal redox reactions in super-Earth and sub-Neptune magma oceans may irreversibly alter the composition of their planetary mantles, and the speciation and longevity of outgassed atmospheres that can be probed with astronomical observations. However, mantle oxidation by iron disproportionation and endogenic production of water from the reaction of primordial H$_2$ atmospheres with mantle FeO hinges on the rate of iron droplet settling to the metal core of sub-Neptunes, and thus the net equilibration rate between atmosphere, mantle, and core. 

Scaling theory predicts that sub-Neptune exoplanets should feature highly turbulent magma oceans because of their higher internal gravity and the depth of the mantle column. The kinetic energy associated with internal fluid circulation can potentially entrain liquid iron droplets in the mantle, which would suspend core formation via droplet settling. This suggests an energetic limit to internal redox reactions that rely on mantle-core equilibration and could enable a composition hysteresis of super-Earths inherited from formation. In the rainout quenched regime reducing conditions in the interior inherited from formation may be sustained, which would result in secondary outgassed atmospheres featuring reduced species.

Iron droplet entrainment from internal magma ocean circulation is sensitive to the planetary heat flow, such that the rainout quenched regime may predominantly operate at larger heliocentric distance or be limited to transient epochs after formation. Vertically resolved models of the coupled evolution of interior, atmosphere, and core of the sub-Neptune to super-Earth transition will be needed to further quantify the relevance of the rainout quenched regime for the speciation of outgassed secondary atmospheres. The resulting deviating compositions from different magma ocean circulation regimes may be imprinted in the atmosphere-stripped sub-population of hot rocky exoplanets, and thus be probed with astronomical observations.

\vspace{1mm}
\small{
\emph{Acknowledgments:} 
I thank Gregor Golabek, Vojt{\v{e}}ch Pato{\v{c}}ka, James Owen, Edwin Kite, and Allona Vazan for discussions and comments that significantly improved clarity and scope of the manuscript. This research was supported by the Simons Foundation (SCOL award No. 611576) and benefitted from information exchange within the program `Alien Earths' (NASA grant No. 80NSSC21K0593) for NASA's Nexus for Exoplanet System Science (NExSS) research coordination network.

\vspace{1mm}
\emph{Software:} \textsc{numpy} \citep{Numpy2020}, \textsc{matplotlib} \citep{Hunter:2007}.
}
\vspace{-1mm}

\bibliography{references}{}
\bibliographystyle{aasjournal}

\end{document}